# Testing Haldane's Conjecture in the $O(3)$ Model by a Meron Cluster Simulation

W. Bietenholz[a], A. Pochinsky[a] [b] and U.-J. Wiese[a] [*]

[a]CTP-LNS, Massachusetts Institute of Technology, Cambridge MA 02139, USA

[b] ITEP, B. Cheremushckinskaya 25, 117259 Moscow, Russia

Twelve years ago, Haldane formulated his famous conjecture for 1-d antiferromagnetic quantum spin chains. In the context of the 2-d $O(3)$ $\sigma$ model with a $\theta$ term, it predicts a phase transition as $\theta \to \pi$, which has not yet been verified reliably. To simulate this we use the Wolff cluster algorithm together with an improved estimator for the charge distribution. Each cluster carries integer or half integer charge. Clusters with charge 1/2 are identified with merons. At $\theta = \pi$ they are inactive, such that the mass gap vanishes. We obtain critical exponents which are consistent with predictions from the $k = 1$ WZNW model, therefore confirming a second order phase transition.

## 1. Haldane's Conjecture and Numerical Results

Antiferromagnetic quantum spin chains are most easily described by the Heisenberg Hamiltonian

$$H = J \sum_{\langle ij \rangle} \vec{S}_i \cdot \vec{S}_j \qquad (1)$$

where $J > 0$ and $\langle ij \rangle$ are nearest neighbors. We deal with quantum spins, therefore

$$[S^a, S^b] = i\epsilon^{abc} S^c \ ; \quad \vec{S}^2 = s(s+1). \qquad (2)$$

This implies that the Néel state, $|s, -s, s \ldots\rangle$, is not an energy eigenstate (for classical spin chains it is the ground state). In fact, the low energy behavior is rather complicated.

In 1983, Haldane conjectured that there is a qualitative difference between integer and half integer $s$ [1]. For half integer spin he predicts a gapless phase, which would have been expected in general from the spin wave theory. The surprising point was that for integer $s$, he conjectured a finite mass gap.

For spin chains, Haldane's conjecture has been proved analytically for all half integer spins [2] and for $s = 1$ [3]. The latter case also agrees with neutron scattering experiments, which found a mass gap in the quasi 1-d $s = 1$ antiferromagnet $CsNiCl_3$ [4], in contrast to $s = 1/2$ systems such as $CuCl_2 \cdot 2\,(NC_5H_5)$. Moreover there is numerical evidence for $s = 2$ [5], but not yet for $2 < s \in \mathbb{Z}$, where the assumed mass gap is tiny.

A low energy effective description (or a large $s$ expansion) leads from the 1-d quantum spin chain to the 2-d $O(3)$ non-linear $\sigma$ model with a $\theta$ vacuum angle [1,6]

$$\theta = 2\pi(s \text{ mod } 1). \qquad (3)$$

Here the poorly understood case flips: while the mass gap at $\theta = 0$ is well known, its vanishing at $\theta = \pi$ has no strong evidence so far. A numerical study by Bhanot et al. did not arrive at conclusive results [7]. Haldane and Affleck have proposed a meron picture to describe this situation. They advocated a second order phase transition at $\theta = \pi$. Affleck extracted quantitative predictions for the critical exponents by an identification with the $k = 1$ WZNW model.

The action of the 2-d $O(3)$ non-linear $\sigma$ model in Euclidean space reads

$$S[\vec{e}] = \int d^2 x [\frac{1}{2g} \partial_\mu \vec{e} \cdot \partial_\mu \vec{e} - \frac{i\theta}{8\pi} \epsilon_{\mu\nu} \vec{e} \cdot (\partial_\mu \vec{e} \times \partial_\nu \vec{e})] \qquad (4)$$

where $|\vec{e}(x)| \equiv 1$ and by large $s$ construction from the antiferromagnetic quantum spin chain one ob-tains $\theta = 2\pi s$. The term $(1/8\pi) \int d^2x \epsilon_{\mu\nu} \vec{e} \cdot (\partial_\mu \vec{e} \times \partial_\nu \vec{e})$ counts how many times $\vec{e}(x)$ covers $S^2$. It is the topological charge $Q \in \mathbb{Z}$. Writing the action as

$$S[\vec{e}] = S_0[\vec{e}] - i\theta Q[\vec{e}] \quad (S_0 \in \mathbb{R}) \qquad (5)$$

we see that physics is $2\pi$ periodic in $\theta$, so we may restrict $\theta$ e.g. to $[0, 2\pi[$ and obtain (3).

[*]Based on a talk presented by W.B. Work supported by U.S. Department of Energy (D.O.E.) under cooperative research agreement DE-FC02-94ER40818.

Haldane and Affleck have introduced the following picture of the mass gap at $\theta = 0$ due to meron condensation [8]. Include a potential $(m^2/2g)\,(e_z)^2$. Thus $\vec{e}$ tends to the $xy$ plane. A vortex in this plane has $Q = 1/2$ and can therefore be called a meron. If the distance $r$ from the origin is infinite, then the vortex should lie strictly in the $xy$ plane, in order to achieve finite action. Close to the origin, however, it has to leave this plane to avoid UV infinite action, in particular $e_z(r = 0) = \pm 1$. Thus merons and antimerons do provide finite contributions to the action and can disorder the system, i.e. cause a mass gap. At $m = \infty$ we have the $O(2)$ model with the Kosterlitz-Thouless phase, understood by vortex condensation. As we decrease $m$ down to 0, the massive phase should cover the whole range of the coupling $g$.

Now we include the topological term. A meron/antimeron picks up the factor $\exp(\pm i\theta/2)$. In particular at $\theta = \pi$ they cancel; they become inactive and can no longer cause a mass gap.

This picture, however, is far from rigorous. Introducing $m$ and sending it to 0 at the end is a dangerous interchange of limits. If we let $m = 0$ from the beginning then the vortex picture does not make any sense at all. Moreover there are multi-vortices which do not cancel at $\theta = \pi$.

In the region addressed by this picture, $\beta = 1/g$ larger than some minimum, we expect a second order phase transition. Since we refer to solid state physics, there is no need for scaling in the continuum limit $\beta \to \infty$. Conformal field theory renders the above picture more quantitative [9].

The $k = 1$ WZNW theory describes a class of models with a massless phase, in particular antiferromagnetic half integer quantum spin chains, the Hubbard model and the $O(3)$ model at $\theta = \pi$. The central charge $k$ must be an integer for the path integral to be well-defined. $k = 1$ is the only case free of relevant operators, therefore it is likely to be a stable fixed point for the above models and it predicts their critical exponents. Marginal operators add logarithmic corrections, which further improve the agreement with our data, see below and [10]. In our case we obtain a mass scale

$$m(\theta) \propto |\pi - \theta|^{2/3} \cdot |\ln(|\pi - \theta|)|^{-1/2}. \qquad (6)$$

Now we can express the topological and magnetic susceptibility $\chi_t$, $\chi_m$ in terms of Fisher's dimensionless parameter $z = Lm$ ($L$: size of the system)

$$\chi_t(\theta) = (L/\sqrt{\ln L})\, g_t(z), \qquad (7)$$
$$\chi_m(\theta) = L\sqrt{\ln L}\, g_m(z). \qquad (8)$$

Close to the phase transition, $g_t$ and $g_m$ are supposed to be universal functions. Our data clearly confirm this. Note that a first order phase transition is ruled out in the regime we simulate, since in that case the susceptibilities would diverge $\propto L^2$. Fig. 1 shows $g_t$, $g_m$ for volumes $V = 18 \times 6$, $24 \times 8$, $30 \times 10$ and $36 \times 12$. We simulated at extremely strong coupling, $\beta = 0$, such that the correlation length at $\theta = 0$ is only 2.8. For each volume we performed at least $10^7$ sweeps. In the plots the measured points are interpolated, without any fitting of free parameters.

## 2. The Algorithm and its Interpretation

The simulation of models with a $\theta$ vacuum term is notoriously difficult. Since a complex Boltzmann factor is not a probability, ordinary Monte Carlo fails. In addition one usually has a sign problem as $\theta$ approaches $\pi$.

We simulate at $\theta = 0$ and then incorporate the factor $\exp(i\theta Q)$. This requires an extremely precise determination of the charge distribution $p(Q)$, which yields the partition function $Z(\theta)$ [11],

$$p(Q) = \int D\vec{e}\, \delta_{Q,Q[\vec{e}]} e^{-S_0[\vec{e}]}, \; Z(\theta) = \sum_Q p(Q) e^{i\theta Q} \quad (9)$$

Observables at fixed charge and $\theta = 0$, and finally expectation values at any $\theta$, are given by

$$\langle \mathcal{O} \rangle_Q = \frac{1}{p(Q)} \int D\vec{e}\, \delta_{Q,Q[\vec{e}]} \mathcal{O}[\vec{e}] e^{-S_0[\vec{e}]},$$

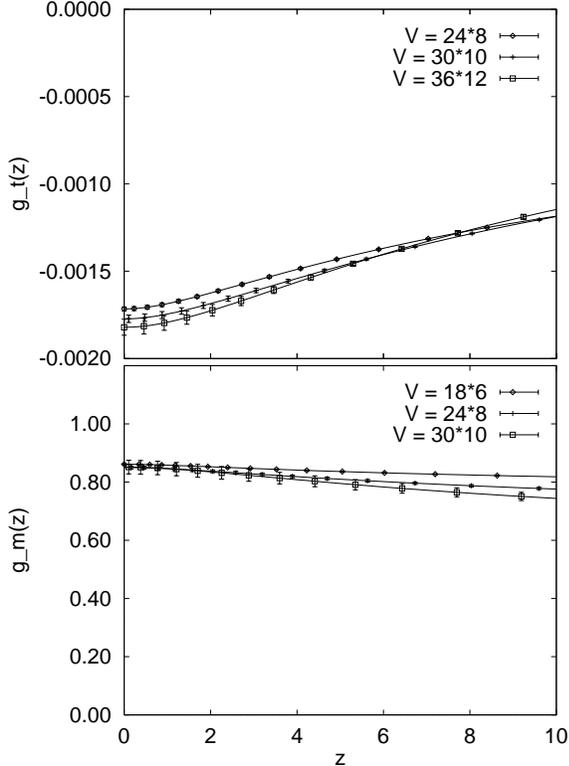

Figure 1. Universal functions $g_t(z)$ and $g_m(z)$ based on numerical data.

$$\langle \mathcal{O} \rangle_\theta = \frac{1}{Z(\theta)} \sum_Q p(Q) \langle \mathcal{O} \rangle_Q e^{i\theta Q}. \quad (10)$$

For technical reasons we deal with a triangular lattice with periodic boundary conditions imposed by a hexagon of side length $L$, such that there are $3L^2$ (independent) sites. We attach to each site a unit vector $\vec{e}_x$. The charge of one triangle is the minimal solid angle spanned by the vectors at its corners. The total charge $Q \in \mathbb{Z}$ is the sum over all triangle charges.

We also specify the lattice action at $\theta = 0$:

$$S_0[\vec{e}] = \sum_{\langle xy \rangle} s(\vec{e}_x, \vec{e}_y)$$

$$s(\vec{e}_x, \vec{e}_y) = \begin{cases} \frac{1}{g}(1 - \vec{e}_x \cdot \vec{e}_y), & \vec{e}_x \cdot \vec{e}_y > -\frac{1}{2} \\ +\infty, & \text{otherwise} \end{cases} \quad (11)$$

i.e. the permitted relative angles of nearest neighbors are $< 2\pi/3$. We are in the universality class of the $\sigma$ model, as we recognize from the symmetries. A priori this lattice action is as good as the standard action. The reason for imposing the constraint is purely technical, see below.

We apply the Wolff cluster algorithm [12]. We first choose a random unit vector $\vec{r}$. 'Flipping' a spin $\vec{e}_x \to \vec{e}_x{}'$ then means to reflect it in the plane through 0 perpendicular to $\vec{r}$. Now we connect neighboring spins $\vec{e}_x$, $\vec{e}_y$ by a bond with probability

$$p = \begin{cases} 0 & \vec{e}_x \cdot \vec{e}_y > \vec{e}_x{}' \cdot \vec{e}_y \\ 1 - e^{s(\vec{e}_x, \vec{e}_y) - s(\vec{e}_x', \vec{e}_y)} & \text{otherwise} \end{cases}.$$

These bonds form clusters. The spins in one cluster can only be flipped collectively, but the clusters are independent. If we flip one cluster, the total charge changes $Q \to Q'$. We define the charge of this cluster as $Q_c = (Q - Q')/2$, an integer or half integer. Now our constraint has the highly non trivial consequence that the charge of one cluster is independent of the orientation of all other clusters. In a typical diagnostics (for $V = 36 \times 12$, $g = \infty$) most clusters are neutral and have a mean size of $\langle n_c \rangle = 1.5$ sites; about 4 percent have $|Q_c| = 1/2$, $\langle n_c \rangle = 10$; 0.1 percent have $|Q_c| = 1$, $\langle n_c \rangle = 32$ and very few clusters have higher charges.

The fact that the cluster charge is determined locally enables the construction of an improved estimator for $p(Q)$ and $\vec{M}^2(Q)$ (the magnetization squared), which are needed for $\chi_t$ and $\chi_m$, respectively. A configuration with $N$ clusters is related to $2^N$ configurations by cluster flips. Simple combinatorics yields the charge distribution for the whole ensemble. Typically we have $O(100)$ clusters, hence the gain in statistics is tremendous. Furthermore $Z(\pi)$ only gets positive contributions from each ensemble, so there is no sign problem in this algorithm. For $\chi_m$ we have to fix $Q$, but still we can flip all the neutral clusters. Thanks to the improved estimator, the improvement factor for the statistics is still huge.

$p(Q)$ varies enormously, but we need to know it precisely over many orders of magnitude. Here we use a trial distribution $p_t(Q)$ (as close as possible to $p(Q)$) together with a reweighting technique [11]. We work with an effective action

$$S_{\text{eff}}[\vec{e}] = S_0[\vec{e}] + \ln p_t(Q[\vec{e}]) \quad (12)$$

related to the charge distribution by

$$p(Q) = p_t(Q) \int D\vec{e} \delta_{Q,Q[\vec{e}]} e^{-S_{\text{eff}}[\vec{e}]}. \qquad (13)$$

A good $p_t(Q)$ makes the charge distribution in $S_{\text{eff}}$ approximately flat. In general it is hard to find, since one does not know $p(Q)$ in advance. But in our case the improved estimator permits the determination of $p(Q)$ to a good accuracy over $\approx 25$ orders of magnitude, as shown in fig. 2. $p_t(Q)$ is then used for flipping the clusters according to the Metropolis algorithm.

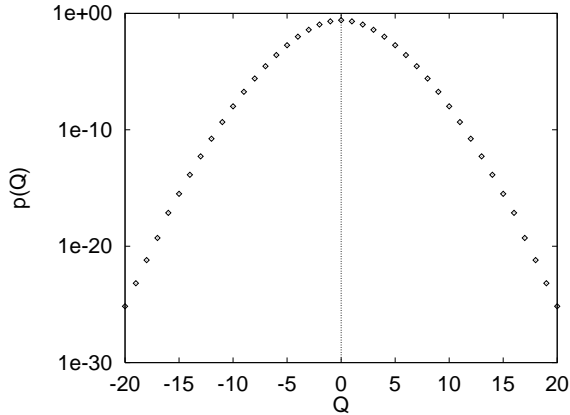

Figure 2. Topological charge distribution

This cluster formulation not only provides a very efficient algorithm, but also suggests a *topological interpretation*. Let us consider the clusters as physical objects, the crucial degrees of freedom. Then clusters with $Q_c = 1/2$ are interpreted as merons. Since the clusters are independent, at $\theta = 0$ the merons build an ideal gas, which causes the mass gap. Now we include the factor $e^{i\theta Q}$; at $\theta = \pi$ it is $(-1)^Q$. If a configuration has at least one cluster of half integer charge, then flipping that cluster changes the sign factor, so the ensemble containing this configuration does not contribute to $Z(\pi)$. The merons and antimerons are 'bound in pairs', they can no longer disorder the system and the mass gap vanishes.

This is a precise formulation of the picture of Haldane and Affleck. The advantage is that it works *without breaking the $O(3)$ symmetry*.

## 3. Outlook

Seiberg has shown that at very strong coupling there is a phase transition of *first order* at $\theta = \pi$ [13]. We simulated at $1/g = \beta = 0$ with the constraint imposed in (11), which corresponds to a $\beta_s > 0$ in the standard lattice action. Since we still observed second order, we could run our algorithm at $\beta < 0$ and try to reach the first order regime. A new study claims agreement with first order, but is plagued by large errors at $\theta \approx \pi$ [14].

It is straightforward to apply our algorithm to the Ising model. This project is in progress.

Finally the topological interpretation of the clusters might even give us a hint for a suitable definition of clusters in lattice gauge theory.

We thank I. Affleck for helpful comments.

## REFERENCES

1. F.D.H. Haldane, Phys. Lett. A93 (1983) 464; Phys. Rev. Lett. 50 (1983) 1153; J. Appl. Phys. 57 (1985) 33.
2. E.H. Lieb, T. Schultz and D. Mattis, Ann. Phys. 16 (1961) 407. I. Affleck and E. H. Lieb, Lett. Math. Phys. 12 (1986) 57.
3. I. Affleck, T. Kennedy, E.H. Lieb and H. Tasaki, Phys. Rev. Lett. 59 (1987) 799; Comm. Math. Phys. 115 (1988) 477.
4. W. Buyers et al., Phys. Rev. Lett. 56 (1986) 371; Phys. Rev. B38 (1988) 543.
5. U. Schollwöck and T. Jolicoeur, Europhysics Lett. 30 (1995) 493.
6. I. Affleck, Nucl. Phys. B257 (1985) 397.
7. G. Bhanot, R. Dashen, N. Seiberg and H. Levine, Phys. Rev. Lett. 53 (1984) 519.
8. F.D.H. Haldane, unpublished. I. Affleck, Phys. Rev. Lett. 56 (1986) 408.
9. I. Affleck in "Fields, String and Critical Phenomena" Les Houches (1988) 563.
10. W. Bietenholz, A. Pochinsky and U.-J. Wiese, Preprint MIT-CTP 2433 (hep-lat/9505019).
11. U.-J. Wiese, Nucl. Phys. B318 (1989) 153.
12. U. Wolff, Phys. Rev. Lett. 62 (1989) 361; Nucl. Phys. B334 (1990) 581.
13. N. Seiberg, Phys. Rev. Lett. 53 (1984) 637.
14. A. Hassan et al., Preprint KYUSHU-HET-26 (hep-lat/9508025).